\documentclass[aps,onecolumn,showpacs,nofootinbib]{revtex4}
\usepackage{graphicx}
\usepackage{epsfig}
\usepackage{amssymb}
\usepackage{graphicx}
\usepackage{amsfonts}
\usepackage{latexsym}
\usepackage{amsmath}
\usepackage{subfigure}
\usepackage{mathrsfs}
\usepackage{float}
\numberwithin{equation}{section}

\newcommand\encadremath[1]{\vbox{\hrule\hbox{\vrule\kern8pt
\vbox{\kern8pt \hbox{$\displaystyle #1$}\kern8pt}
\kern8pt\vrule}\hrule}}
\def\enca#1{\vbox{\hrule\hbox{
\vrule\kern8pt\vbox{\kern8pt \hbox{$\displaystyle #1$} \kern8pt}
\kern8pt\vrule}\hrule}}

\begin{document}
\title{Generalised Freud's equation and level densities with polynomial potential}
\author{Akshat Boobna}\email{akshatb42@gmail.com}
\affiliation{The Creative School, E-791, C.R. Park, New Delhi-110017}
\author{Saugata Ghosh}\email{saugata135@yahoo.com}
\affiliation{The Creative School, E-791, C.R. Park, New Delhi-110017 }
\date{\today}
\begin{abstract}
We study orthogonal polynomials with weight $\exp[-NV(x)]$, where $V(x)=\sum_{k=1}^{d}a_{2k}x^{2k}/2k$ is a polynomial of order $2d$. We derive the generalised Freud's equations for $d=3$, $4$ and $5$ and using this obtain $R_{\mu}=h_{\mu}/h_{\mu -1}$, where $h_{\mu}$ is the normalization constant for the corresponding orthogonal polynomials. Moments of the density functions, expressed in terms of $R_{\mu}$, are obtained using Freud's equation and using this, explicit results of level densities as $N\rightarrow\infty$ are derived.
\end{abstract}
\maketitle

\section{Introduction}


Universality in random matrix theory \cite{beenakker,deift2,deift3,deift4} has led people to study  orthogonal \cite{szego,deift1,She} and skew-orthogonal polynomials \cite{ghoshb} in great details. However, in the process, the non-universal level densities are neglected inspite of the possibility of its direct application in various physical systems. In this context, we study level densities of a class of non-Gaussian random matrix ensembles and thereby develop the theory of orthogonal polynomials.

Orthogonal polynomials are defined as
\begin{equation}
\int_{\mathbb{R}}P_{n}(x)P_{m}(x)w(x)dx=h_{n}\delta_{nm},\hspace{1cm}n,m\in\mathbb{N}.\label{ortho}
\end{equation}

We study orthogonal polynomials with weight function $w(x)=\exp(-NV(x))$, where
\begin{equation}
\label{vx:gen}
V(x)=\sum_{k=1}^{d}a_{2k}x^{2k}/(2k),\hspace{1cm} a_{2d}>0.
\end{equation}
Here, we make a numerical analysis of  orthogonal polynomials corresponding to $d=3$, $4$ and $5$.
We derive the corresponding Freud's equation and calculate $R_{\mu}=h_{\mu}/h_{\mu-1}$. We observe interesting patterns in the behavior of $R_{\mu}$.

Once we have an understanding of $R_{\mu}$, we use these results to obtain level densities of non-Gaussian ensembles of random matrices. We know that variation of the first n-eigenvalues of a random matrix can be studied by the n-point correlation function, $R_{n}^{(\beta)}(x_1,...,x_n)$ which is defined by
\begin{equation}
R_{n}^{(\beta)}(x_1,...,x_n)=\frac{N!}{(N-n)!}\int_{\mathbb{R}^{N-n}}dx_{n+1}...dx_{N}P_{\beta,N}(x_1,...,x_N),
\end{equation}
where $\beta=1,2,4$ correspond to ensembles of random matrices invariant under orthogonal, unitary and symplectic transformations. This allows us to find the probability density of the $n$ eigenvalues at $x_1,...,x_n$, irrespective of the eigenvalues at $x_{n+1}...x_{N}$. $R_{1}^{(\beta)}(x)$ is the level density, which, for $\beta=2$ can be written as \cite{dyson1,dysommehta,mehta,mehta1}
\begin{equation}
\label{den1}
R_1^{(2)}(x)=\sum_{\mu=0}^{N-1}(h_{\mu})^{-1}[P_{\mu}(x)]^2 e^{-NV(x)}.
\end{equation}
To calculate $R_{1}^{(2)}(x)$ as $N\rightarrow\infty$, the standard method is to use the Christoffel Darboux formula and the asymptotic results
of orthogonal polynomials. The latter is not always available for general polynomial potential  inspite of some serious contributions from several  authors \cite{plancherel,deift1,deift5,deift6,deift7} on the asymptotics of orthogonal polynomials with $V(x)=x^{2d}$ using the Riemann Hilbert technique. In this paper, we use the method of resolvent to obtain the level densities as $N\rightarrow\infty$.This needs an understanding of  moments $M_{k}$ defined as
\begin{equation}
\label{mom}
M_{k}=\int_{\mathbb{R}} x^{k}R_1^{(2)}(x)dx,\hspace{1cm}k\in\mathbb{N}.
\end{equation}
This is derived using the values of $R_{\mu}$ using generalised Freud's equation, which we derive independently. Using this, we obtain the corresponding level densities. This gives us a good understanding of the origin of multiple band formation in the level densities in polynomial potential.

The paper is organized as follows: In section 2, we study the $d=3$ case and observe the behaviour of $R_{\mu}$ for different values of $a_k$. Section 3 and 4 deal with $d=4$ and $d=5$ results. This is followed by our concluding remarks.


\section{$d=3$ case}


\subsection{Freud's equation}
Orthogonal monic polynomials with even weight satisfy a recursion relation \cite{szego}
\begin{equation}
xP_{\mu}=P_{\mu+1}+R_{\mu}P_{\mu-1}, \hspace{2cm} \mu\in \mathbb{N}, \label{3term}
\end{equation}
where $R_{\mu}=h_{\mu}/h_{\mu-1}$, for $\mu\geq1$ and $R_{0}=0$.

A major development in the study of quartic weight ($d=2$ in eq.(\ref{vx:gen})) polynomials \cite{stoj1,stoj1a,stoj2} was the following recursive equation in $R_{\mu}$ due to \cite{Freud}.
\begin{equation}
\mu+1=NR_{\mu+1}[a_4 (R_{\mu+2}+R_{\mu+1}+R_{\mu})+a_2].
\end{equation}
Now, we derive a similar Freud's equation for sextic potential, i.e. $d=3$ in eq.(\ref{vx:gen}). We use the identity
\begin{equation}
\int dx[P_{\mu+1}(x)P_{\mu}(x)e^{-NV(x)}]'=0.
\end{equation}
Using $P_{\mu}(x)=x^{\mu}+\ldots$ and the orthonormality condition (\ref{ortho}), we get
\begin{equation}
\int [e^{-NV(x)}][P'_{\mu+1}(x)P_{\mu}(x)+P_{\mu+1}(x)P'_{\mu}] dx+\int N[a_6 x^5+ a_4 x^3+ a_2 x]P_{\mu+1}(x)P_{\mu}(x)e^{-NV(x)}] dx =0 \nonumber \\
\end{equation}
This gives us
\begin{equation}
(\mu+1)h_{\mu}=\int N a_6 x^5P_{\mu+1}(x)P_{\mu}(x)e^{-NV(x)}dx] + \int N a_4 x^3P_{\mu+1}(x)P_{\mu}(x)e^{-NV(x)}dx] + \int N a_2 xP_{\mu+1}(x)P_{\mu}(x)e^{-NV(x)}dx]. \nonumber
\end{equation}
Using (\ref{3term}), we obtain
\begin{equation}
\begin{split}
\mu+1=NR_{\mu+1}[&a_6(R_{\mu+2}(R_{\mu}+R_{\mu+1}+R_{\mu+2}+R_{\mu+3})\\
&+R_{\mu+1}(R_{\mu}+R_{\mu+1}+R_{\mu+2})\\
&+R_{\mu}(R_{\mu-1}+R_{\mu}+R_{\mu+1}))\\
   & +a_4(R_{\mu-1}+R_{\mu}+R_{\mu+1})+a_2].
\end{split}
\label{freud3}
\end{equation}
Here we note that the corresponding Freud's equation is cubic in nature thereby giving rise to oscillatory behavior.

\begin{figure}[h]
        \includegraphics[scale=1]{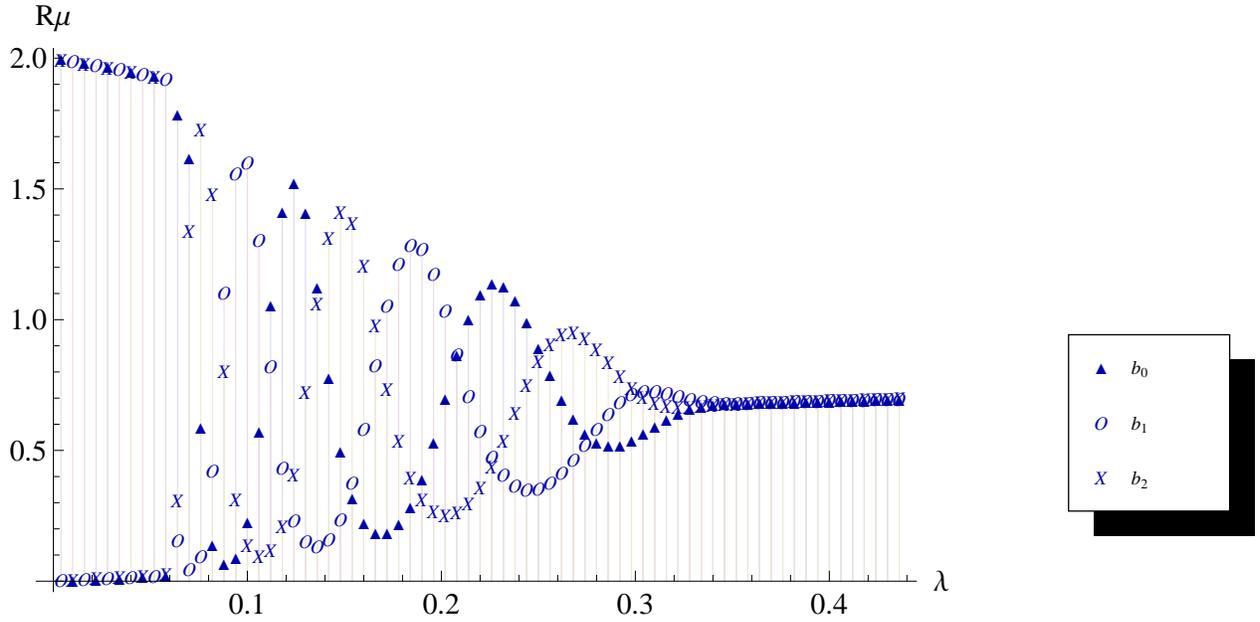}
        \caption{$R_{\mu}$ plot modulo 3 for $d=3$,$a_6=1$, $a_4=-2.5$, $a_2=1$, $N=500$ using eq. (\ref{freud3})}\label{fig:sextic}
\end{figure}

\subsection{The R$\mu$ plot}

For the $d=2$ case, two main features were observed in the $R_{\mu}$ plot from the original Freud's equation: A two band structure formed by an oscillation between two values, converging to a single band.

In the sextic case, the two band and single band structures reappear, however a new, more chaotic structure is also seen, appearing either between one band and one band or one band and two band structures. Henceforth, it is termed as a "transient structure".
\subsubsection{Single band structure}
For the single band structure, all the terms in the Freud's equation are equal to each other. Thus, we obtain a $\lambda$ dependent cubic equation (where $\lambda=\mu/N$) , solving which we obtain one real solution
\begin{equation}
D_{\pm}=\frac{27}{2}[(2a_4^3-10a_6a_4-100a_6^2\lambda)\pm\sqrt{(2a_4^3-10a_6a_4-100a_6^2\lambda)^2 - 4(a_6^2-\frac{10a_6a_2}{3})^3}
\end{equation}
\begin{equation}
R_{\mu}=-\frac{a_4}{10a_6}-\frac{\sqrt[3]{D_{+}}}{30a_6}-\frac{\sqrt[3]{D_{-}}}{30a_6}
\end{equation}
which gives the value of $R_{\mu}$ for $\lambda$ values where single band exists.
\subsubsection{Two band structure}
Solving the Freud's equation assuming that two bands are formed (as seen), i.e. \\
$A_0=R_0=R_2=R_4=...$ \\
$A_1=R_1=R_3=R_5=...$ \\
for $N\gg\mu$, we get
\begin{equation}
A_0+A_1=\frac{-a_4+\sqrt{a_4^2-4a_2a_6}}{2a_6}
\end{equation}
It has been numerically verified that the bottom band (A1) tends to $0$, and we find that $A_1\propto 1/N$ and $A_1\propto 1/(a_4)^2$

\subsubsection{Transient structure}

When the transient structure is divided modulo 3 into three bands ($b_0$, $b_1$ and $b_2$ in fig. \ref{fig:sextic}), it is seen that each of the three separate bands continuously oscillate, and converge to a common value.

The sum mod 3 for the duration of the transient structure oscillates above the value
\begin{equation}
b_0+b_1+b_2\approx \frac{-a_4+\sqrt{a_4^2-4a_2a_6}}{2a_6},
\end{equation}
where $b_0$, $b_1$ and $b_2$ correspond to three consecutive values from each of the bands.
\subsection{Critical $a_4$'s}
On analyzing the roots of the sextic potential, we obtain the critical value of $a_4$, denoted here by $a_{4c}$, where the structure of the $R_{\mu}$ plot changes. We find the points at which potential plot touches $0$, and obtain
\begin{equation}
a_{4c}=-\sqrt{\frac{48}{9}a_2 a_6}
\end{equation}

In the $R_{\mu}$ plot, when
(i) $a_4 <a_{4c}$, we observe a two band structure, followed by a transient structure, which converges to single band.
(ii)$ a_4 > a_{4c}$, we see a one band structure, followed by a transient structure, which converges to single band.

\begin{figure} [H]
        \subfigure[$a_4=-2.25$]{
                \includegraphics[scale=0.7]{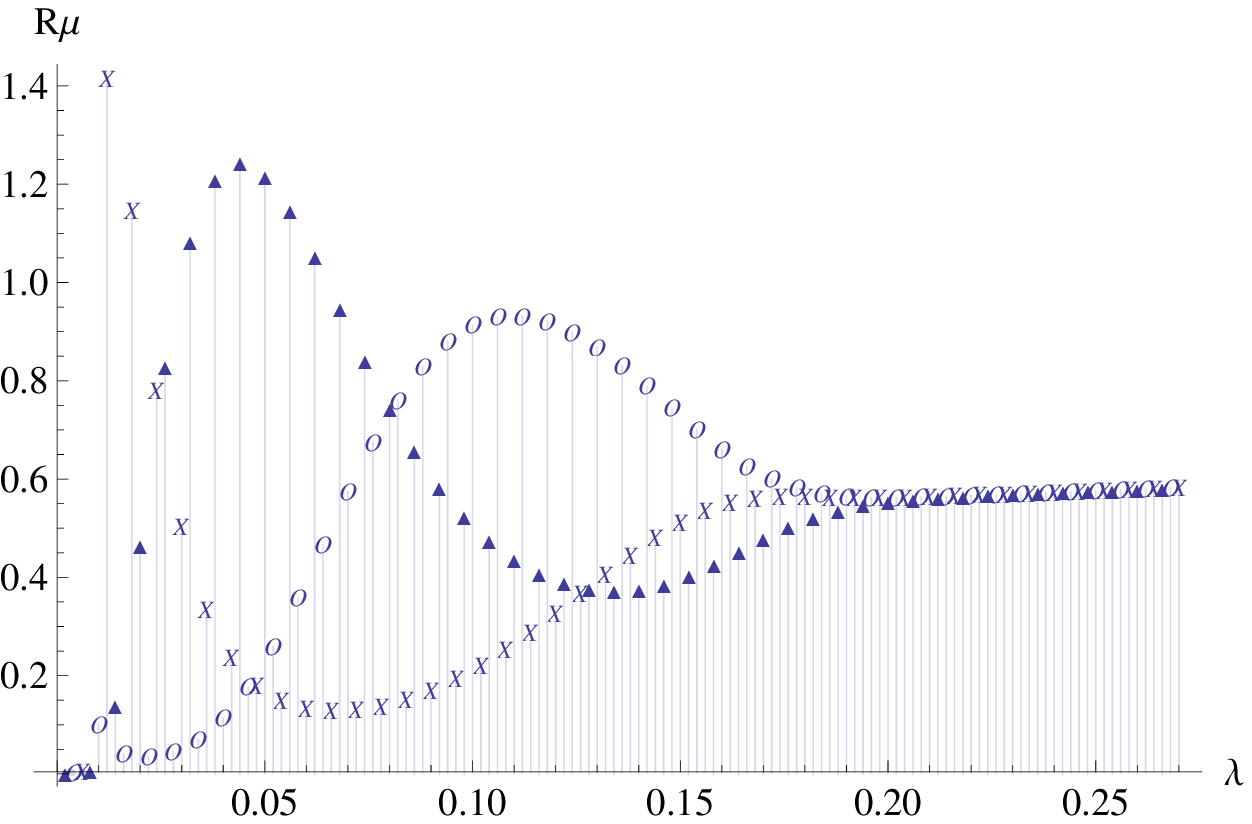}
        }
       \subfigure[$a_4=-2.30$]{
                \includegraphics[scale=0.7]{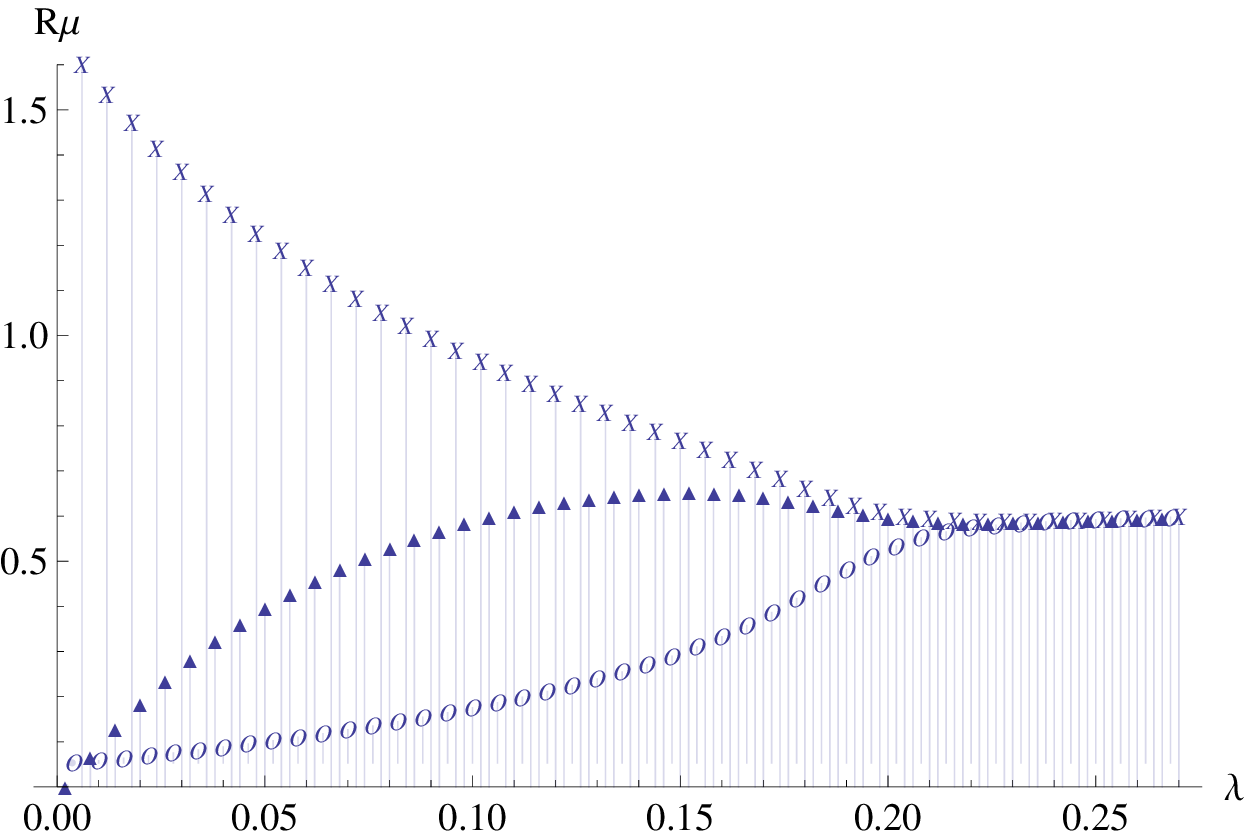}
        }

        \subfigure[$a_4=-\sqrt{48/9}$ ($=a_{4c}$)]{
                \includegraphics[scale=0.7]{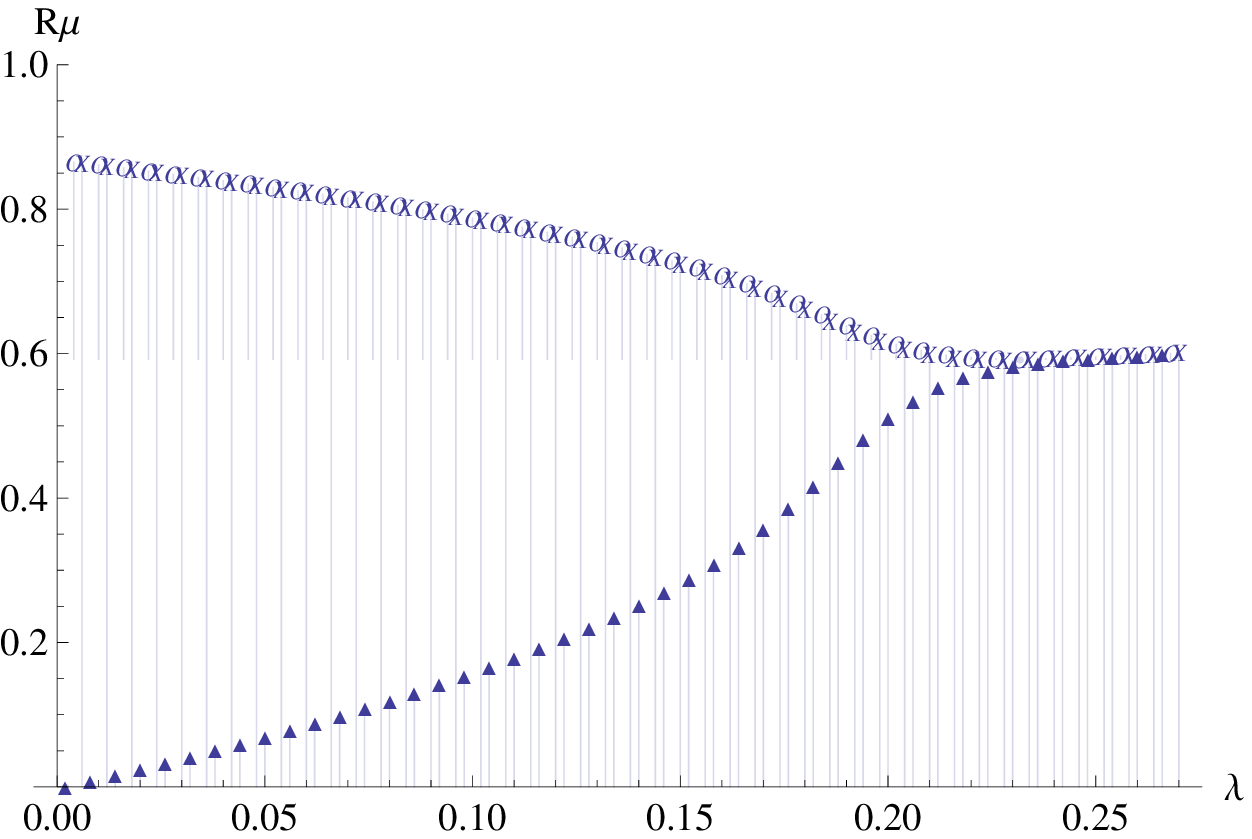}
        }
        \subfigure[$a_4=-2.35$]{
                \includegraphics[scale=0.7]{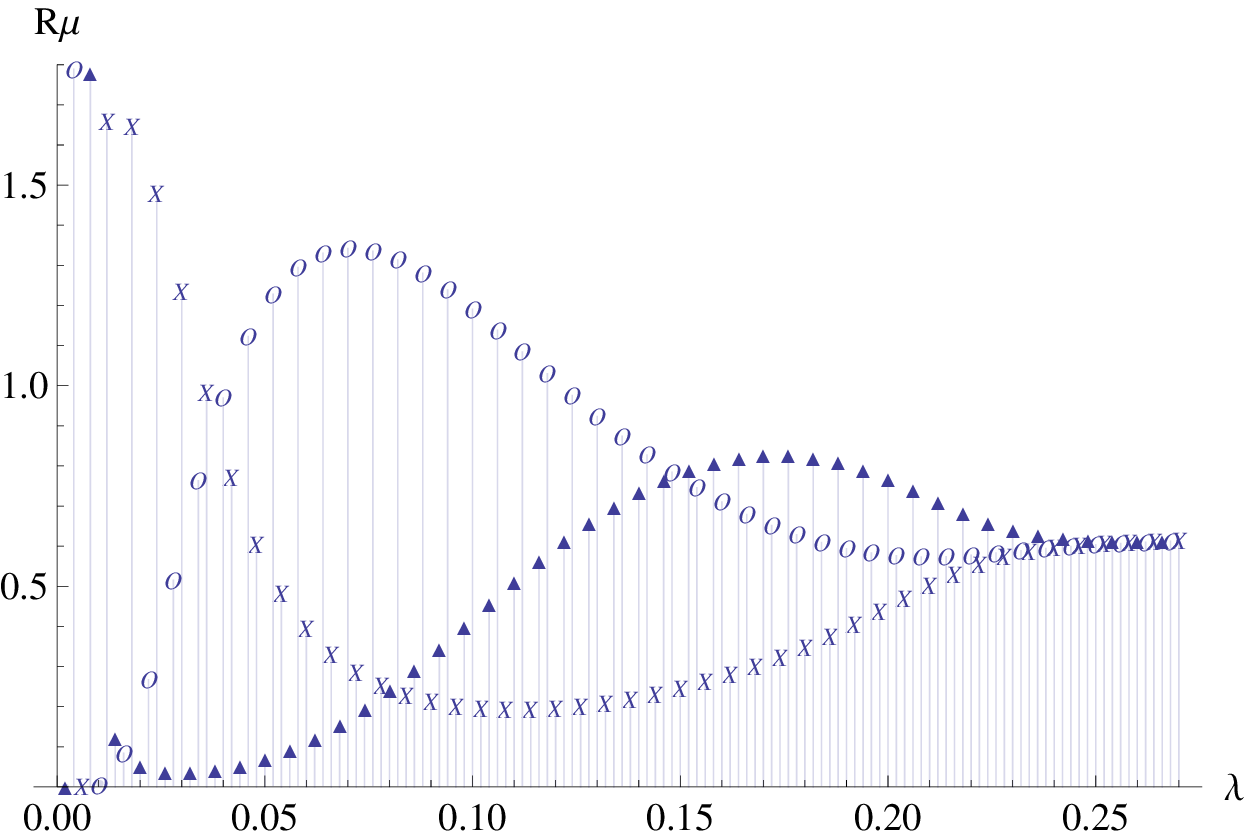}
        }
        \caption{$R_{\mu}$ plot as $a_4$ approaches $a_{4c}$ for $a_6$=1, $a_2$=1, N=500}
        \label{critical}
\end{figure}

We now analyze the behavior of $R_{\mu}$ as $a_4$ approaches $a_{4c}$ (fig. \ref{critical}). It is observed the transient structure resolves into three distinct bands near $a_{4c}$. Exactly at the critical value, two of these bands coincide to form an upper band, and the third band forms a lower band, creating a pseudo two band structure.

\subsection{Level Density}\label{secld}
In this subsection, we derive the level density using the method of resolvent \cite{ghosh,ghoshb} as $N\rightarrow\infty$. The result is expressed in terms of moments $M_{k}$ (\ref{mom}) which are derived using results from the Freud's equation. We then compare the result with the level density for $N=30$ using (\ref{den1}).

\begin{figure}[h]
        \includegraphics[scale=0.6]{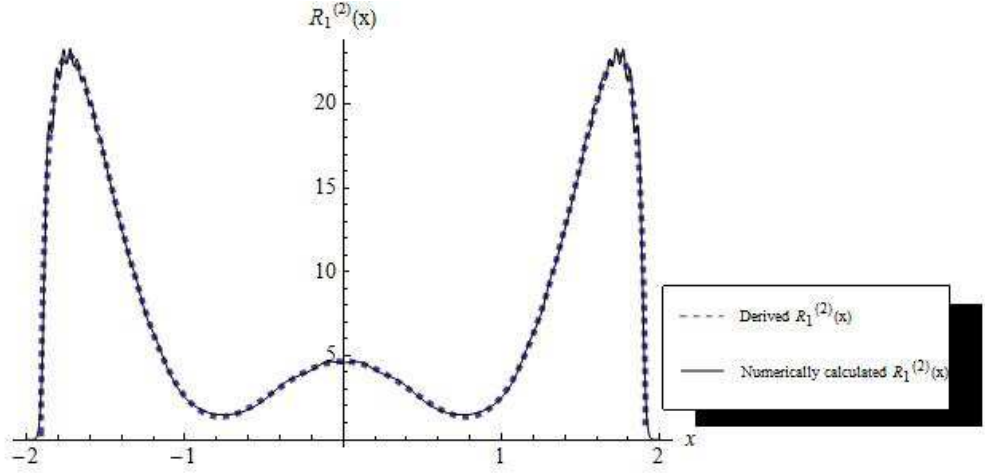}
        \caption{Level density plots for $d=3$, $a_6=1$, $a_4=-3$, $a_2=1$ $N=30$.
        For the theoretical plot, calculated moments are $M_2=62.536$, $M_4=164.770$.}
        \label{ld}
\end{figure}
Here, we recall \cite{ghoshb}that we use the scaling
$V(x)\rightarrow V(x)/2$ to obtain the corresponding results.

\begin{eqnarray}
{[\pi R_{1}^{(2)}(x)]}^{2} &=& N\int_{-\infty}^{\infty} \frac{V'(z)-V'(x)}{z-x}R_{1}^{(2)}(x)dx - N^2 \left[\frac{V'(x)}{2}\right]^2 \nonumber \\
&=& N \left[a_6(x^4N+x^2M_2+M_4) + a_4 (x^2N+M_2) +a_2N \right] - N^2 x^2 \left(\frac{a_6x^4+ a_4x^2+ a_2}{2}\right)^2. \nonumber
\end{eqnarray}
Finally,
\begin{equation}
\left[{\frac{\pi R_1^{(2)}(x)}{N}}\right]^{2}=\left(\frac{a_6M_{4}+a_4M_{2}}{N}+a_2\right)+x^2\left[\left(a_6x^2+\frac{a_6M_{2}}{N}
+a_4\right)-\frac{1}{4}\left(a_6x^4+a_4x^2+a_2\right)^2\right].
\end{equation}

Now that we have derived $R_{1}^{(2)}(x)$ in terms of $M_{2}$ and $M_{4}$, we would be interested to calculate them using Freud's equation.
We use
\begin{eqnarray}
\nonumber
M_{k} &=&\sum_{\mu}\int\frac{x^{k}P_{\mu}^{2}(x)}{h_{\mu}}w(x)dx\\
\nonumber
      &=&\sum_{\mu}\int\frac{(x^{k}P_{\mu}(x))P_{\mu}(x)}{h_{\mu}}w(x)dx\\
\nonumber
      &=&\sum_{\mu}\sum_{\nu}\int\frac{C_{\nu}P_{\nu}(x)P_{\mu}(x)}{h_{\mu}}w(x)dx\\
      &=&\sum_{\mu}C_{\mu},
\end{eqnarray}
where $C_{\mu}$ are coefficients which can be expressed in terms of $R_{\mu}$. A few typical examples are

\begin{equation}
M_2=\sum_{\mu=0}^{N-1}
\left(R_{\mu +1}+R_{\mu}\right),
\label{M2}
\end{equation}

\begin{equation}
M_4=\sum_{\mu=0}^{N}
\left(R_{\mu}^2+R_{\mu+1}^2+2R_{\mu}R_{\mu+1}+R_{\mu+1}R_{\mu+2}+R_{\mu}R_{\mu-1}\right),
\label{M4}
\end{equation}

and

\begin{eqnarray}
\nonumber
M_6 = \sum_{\mu=0}^{N} &&(R_{\mu-2} R_{\mu-1} R_{\mu}+R_{\mu-1}^2 R_{\mu}+2
R_{\mu-1} R_{\mu}^2+R_{\mu}^3+2 R_{\mu-1} R_{\mu} R_{\mu+1}\\&&
\nonumber
+3R_{\mu}^2 R_{\mu+1}+3 R_{\mu} R_{\mu+1}^2+R_{\mu+1}^3+2 R_{\mu}
R_{\mu+1} R_{\mu+2}+2 R_{\mu+1}^2 R_{\mu+2}\\&&
+R_{\mu+1}R_{\mu+2}^2+R_{\mu+1} R_{\mu+2} R_{\mu+3}).
\label{M6}
\end{eqnarray}

The expression for $M_8$ and higher moments are extremely cumbersome but can be easily calculated using the aforementioned algorithm.
\section{$d=4$ case}

\subsection{Freud's equation}

\begin{figure}[h]
    \includegraphics[scale=0.9]{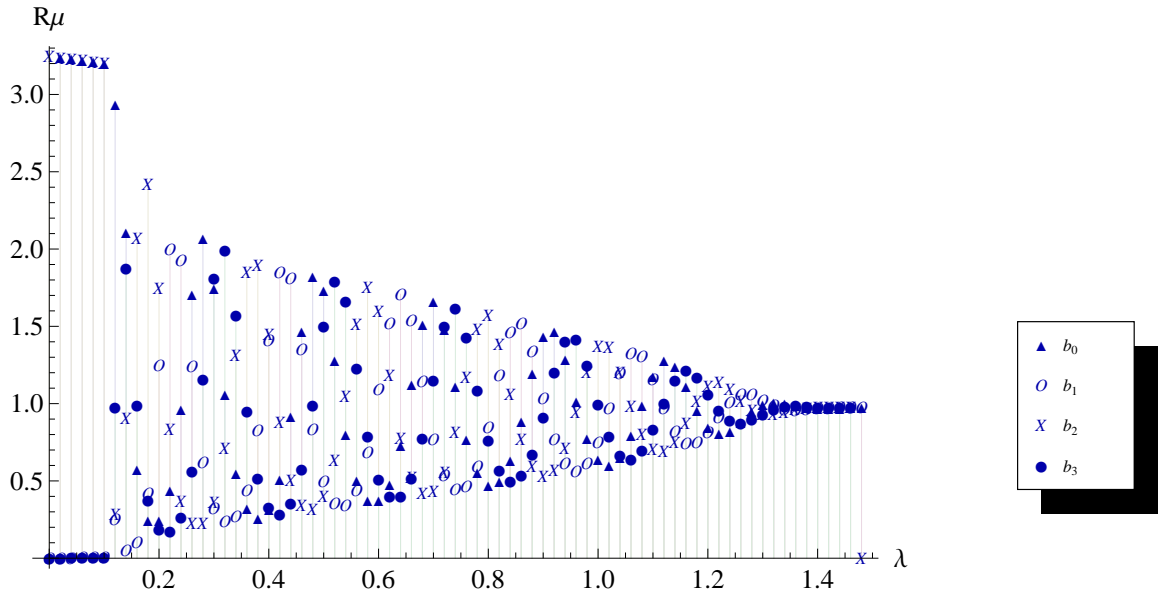}
    \caption{$d=4$ case plot for $a_8=1$, $a_6=-5$, $a_4=6$, $a_2=-1$, $N=200$}\label{octic}
\end{figure}

As in the $d=3$ case, we start with the identity
\begin{equation}
\int dx[P_{\mu+1}(x)P_{\mu}(x)e^{-NV(x)}]'=0,
\end{equation}
 where $V(x)$ is defined as in eq.(\ref{vx:gen}), but with $d=4$. Using the recursion relation for orthogonal polynomials, we get
 \begin{equation}
\begin{split}
\mu+1=NR_{\mu+1}[&a_8(R_{\mu+2}R_{\mu+3}\sum_{i=\mu}^{\mu+4}R_i + R_{\mu+2}^2\sum_{i=\mu}^{\mu+3}R_i+R_{\mu+2}R_{\mu+1}\sum_{i=\mu}^{\mu+2}R_i \\
&+ R_{\mu+2}R_{\mu}\sum_{i=\mu-1}^{\mu+1}R_i +R_{\mu+1}R_{\mu+2}\sum_{i=\mu}^{\mu+3}R_i+R_{\mu+1}^2\sum_{i=\mu}^{\mu+2}R_i +R_{\mu}R_{\mu+1}\sum_{i=\mu-1}^{\mu+1}R_i\\
&+R_{\mu}R_{\mu+1}\sum_{i=\mu}^{\mu+2}R_i+R_{\mu}R_{\mu-1}\sum_{i=\mu-2}^{\mu+1}R_i +R_{\mu}^2\sum_{i=\mu-1}^{\mu+1}R_i) \\
&+a_6(R_{\mu+2}\sum_{i=\mu}^{\mu+3}R_i+R_{\mu+1}\sum_{i=\mu}^{\mu+2}R_i+R_{\mu}\sum_{i=\mu-1}^{\mu+1}R_i)\\
   & +a_4\sum_{i=\mu-1}^{\mu+1}R_i+a_2]
\end{split}
\end{equation}

Here we note that due to the non-linear nature of the Freud's equation, we observe oscillations in the solution for $R_{\mu}$. These oscillations can be seen when the plot is divided modulo 4 into 4 residual bands ($b_0$,$b_1$,$b_2$ and $b_3$ in fig. \ref{octic})

\subsection{Level density}
\begin{figure}[h]
        \includegraphics[scale=0.6]{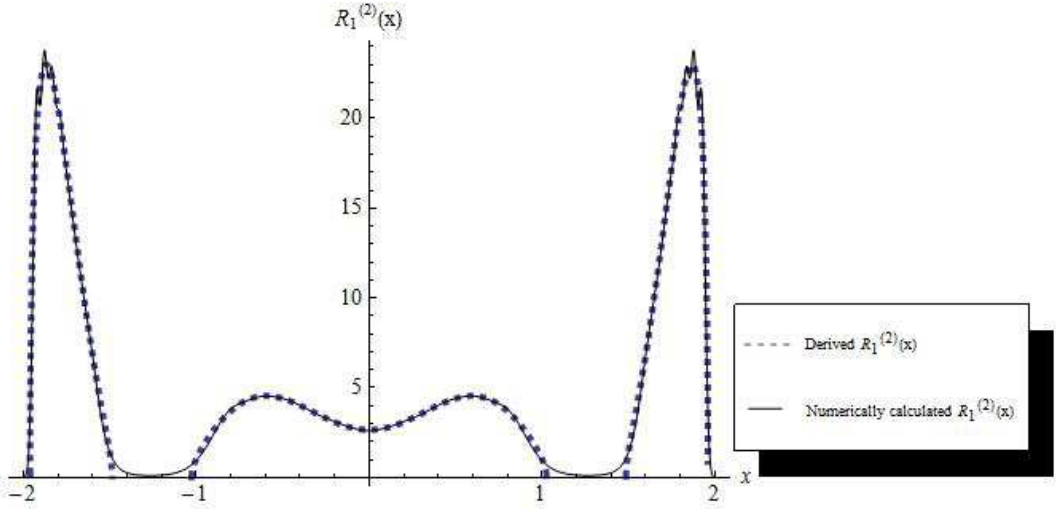}
        \caption{Level density plots for $d=4$, $a_8=1$, $a_6=-5$, $a_4=6$, $a_2=-1$, $N=20$.
        For the theoretical plot, calculated moments are $M_2=43.475$, $M_4=134.555$, $M_6=438.400$.}
        \label{sld}
\end{figure}

Using the obtained moments and the formulation for finding level density (sec. \ref{secld}), we derive the function for $R_1^{(2)}(x)$ for the $d=4$ case (\ref{eq:ld4}). The plot of $R_1^{(2)}(x)$ obtained using this function is compared with the $R_1^{(2)}(x)$ calculated from (\ref{den1}) in fig. \ref{sld}.

We note that irregularities in the form of small oscillations around the expected value are seen near the peaks. This is because the numerically calculated level density is for a finite value of $N$. These oscillations gradually disappear as the value of $N$ increases.

\begin{eqnarray}
\nonumber
\left[{\frac{\pi R_1^{(2)}(x)}{N}}\right]^{2}&=& \left[\frac{a_8 M_{6}+a_6 M_{4}+a_4 M_{2}}{N}+a_2\right]
\\ &&+x^2\left(a_8 x^4+a_6 x^2+a_4+\frac{a_8 x^2M_{2}+a_8 M_{4}+a_6 M_{2}}{N}-\frac{(a_8 x^6+a_6 x^4+a_4 x^2+a_2)^2}{4}\right).
\label{eq:ld4}
\end{eqnarray}

\section{$d=5$ case}

\subsection{Freud's equation}
As in the $d=3$ case, we start with the identity
\begin{equation}
\int dx[P_{\mu+1}(x)P_{\mu}(x)e^{-NV(x)}]'=0,
\end{equation}
 where $V(x)$ is defined as in eq.(\ref{vx:gen}), but with $d=5$.
From here, one can understand that finding the Freud's equation is algorithmic in nature and we leave it to the reader to derive it explicitly. Here, we
will show the generic plot of the $R_{\mu}$ function.

\begin{figure}[h]
        \includegraphics[scale=1]{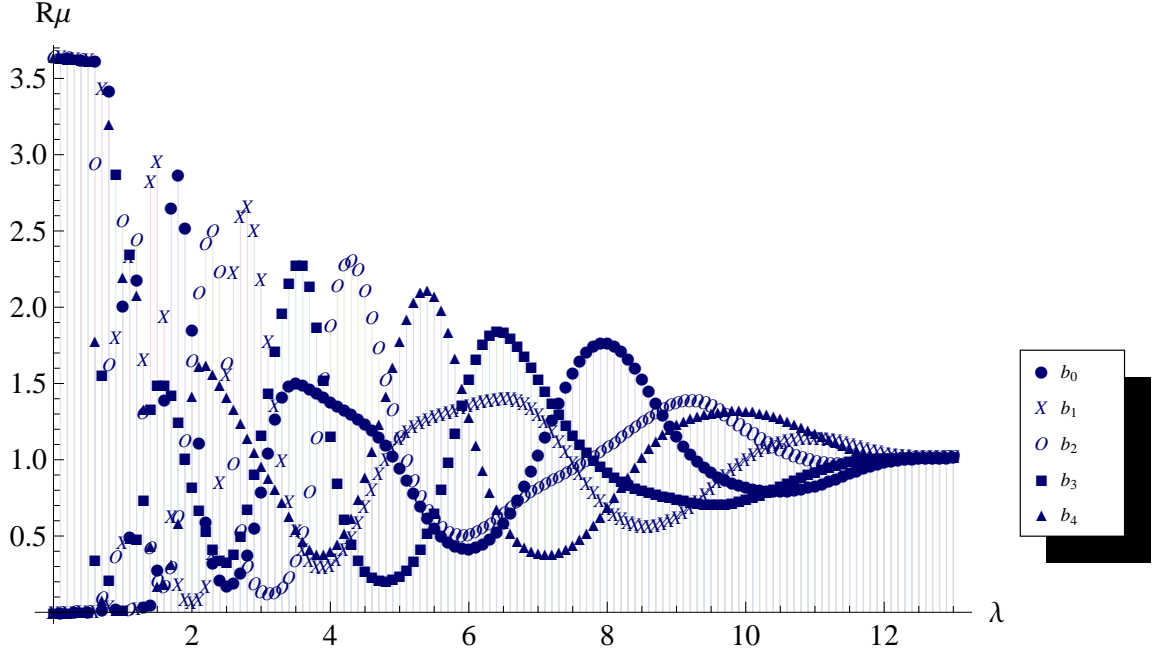}
        \label{fig:decic_osc}
        \caption{$d=5$ case plot modulo 5 for $a_{10}=10$, $a_8=-80$, $a_6=210$, $a_4=-200$, $a_2=48$, $N=50$}
\end{figure}
\subsection{Level density}

\begin{figure}[h]
        \includegraphics[scale=0.65]{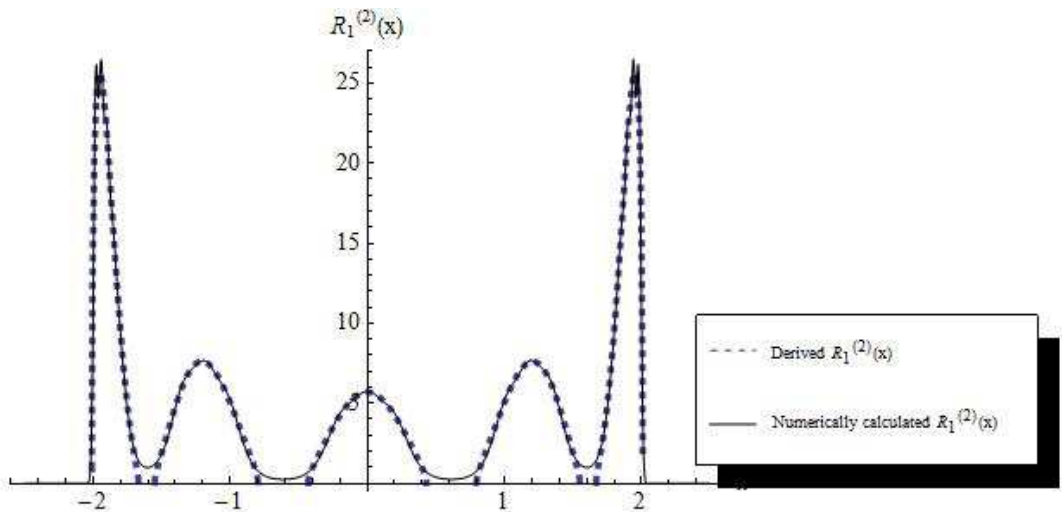}
        \caption{Level density plot for for $d=5$, $a_{10}=1$, $a_8=-8$, $a_6=21$, $a_4=-20$, $a_2=4.8$, $N=20$. For the theoretical plot, calculated moments are $M_2=43.960$, $M_4=136.008$, $M_6=460.375$, $M_8=1625.995$.}
        \label{fig:decicld}
\end{figure}

Having derived the moments $M_2$, $M_4$, $M_6$ and $M_8$ using the general formulation given in sec. \ref{secld}, we use the derivation provided to obtain the function for $R_1^{(2)}(x)$ for the $d=5$ case (\ref{eq:ld5}). The plot of $R_1^{(2)}(x)$ obtained using this function is compared with the $R_1^{(2)}(x)$ calculated from (\ref{den1}) in fig. \ref{fig:decicld}.

Once again, we note that small oscillations around the expected value are seen. This is because we are calculating $R_1^{(2)}(x)$ for a finite value of $N$, and these become smooth as $N\rightarrow\infty$
\begin{equation}
\begin{split}
\left[{\frac{\pi R_1^{(2)}(x)}{N}}\right]^{2}=&\frac{a_{10}M_8+a_8M_6+a_6M_4+a_4M_2}{N}+a_2 \\
&+x^2\left(a_{10}x^6+a_8x^4+a_6x^2+a_4+\frac{a_{10}x^4M_2+a_8x^2M_2+a_6M_2+a_{10}x^2M_4+a_{10}M_6+a_8M_4}{N}\right) \\
&-\frac{x^2}{4}(a_{10}x^8+a_8x^6+a_6x^4+a_4x^2+a_2)^2.
\end{split}
\label{eq:ld5}
\end{equation}

\section{Conclusion}

In this paper, we obtain the Freud's equation for polynomials with weight function $\exp[-NV(x)]$, where $V(x)=\sum_{k=1}^{d}a_{2k}x^{2k}/2k$ is a polynomial of order $2d$. We derive the generalised Freud's equations for $d=3$, $4$ and $5$. We observe limit cycle behavior of $R_{\mu}$. We use these results and the method of resolvent to obtain the level densities of the
corresponding random matrix models. However, this involves an explicit calculation of the higher moments which we calculate numerically and insert in the analytic results of the level density. It would be nice to obtain explicit results of these moments as was done for the quartic case. But we have failed in this investigation due to the complex nature of the Freud's equation.

Here, we might recall that for $d=2$, the Freud's equation is quadratic while for higher $d$, it becomes cubic ($d=3$), quartic ($d=4$) and so on. This results in oscillations in the $R_{\mu}$ function and hence studying the limit cycle behavior becomes increasingly complicated. Further investigation is needed to study these non-linear Freud's
equations, specially in the context of integrability and hence the existence of Lax pairs. We wish to come back to these questions in a later publication.

\end{document}